\documentclass[aps,amssymb,amsmath,12pt,nofootinbib]{revtex4}
\usepackage{epsfig}

\begin{document}

\def\bigO{\mathcal{O}}
\def\be{\begin{equation}}
\def\ee{\end{equation}}
\def\lb{\label}
\def\Sig{\Sigma}
\def\Sigp{\Sigma_{+}}
\def\Sigm{\Sigma_{-}}
\def\Sigc{\Sigma_{\times}}
\def\Sigt{\Sigma_2}
\def\Np{N_{+}}
\def\Nm{N_{-}} 
\def\Nc{N_{\times}}
\def\EEE{E_1{}^1}
\def\hatN{\hat{N}}
\def\hatS{\hat{\Sigma}}
\def\hatEEE{\hat{E}_1{}^1}

\title{Asymptotic analysis of spatially inhomogeneous 
		stiff and ultra-stiff cosmologies}
\author{A. A. Coley}
\email{aac@mathstat.dal.ca}
\affiliation{Department of Mathematics and Statistics,
Dalhousie University\\
Halifax, Nova Scotia, Canada \enskip B3H 3J5}
\author{W. C. Lim}
\email{wclim@mathstat.dal.ca}
\affiliation{Department of Mathematics and Statistics,
Dalhousie University\\ 
Halifax, Nova Scotia, Canada \enskip B3H 3J5} 

\begin{abstract}

We calculate analytically the past asymptotic decay rates close to an
initial singularity in general $G_{0}$ spatially inhomogeneous perfect
fluid models with
an effective equation of state which is stiff or ultra-stiff (i.e.,
$\gamma
\ge 2$). These results are then supported by numerical simulations
in a special class of $G_{2}$ cosmological models. Our analysis confirms
and extends the BKL conjectures and lends support to recent isotropization
results in cosmological models of current interest (with $\gamma > 2$).

\end{abstract}

\maketitle

\section{Introduction}

The conjectures of Belinski\v{\i}, Khalatnikov and Lifshitz (BKL)
\cite{BKL}
assert that the structure of space-time singularities in
general relativity (GR) have the following properties 
(for a perfect fluid with $\gamma\le 2$, where $(\gamma-1)\equiv p/\rho$ is defined as the ratio of
the pressure $p$ to the energy density $\rho$):
1. Each spatial point evolves towards the singularity as
if it were a spatially homogeneous cosmology.
2. Space-times with
non-stiff matter, $\gamma<2 $, have
the property that asymptotically
close to the singularity matter is not dynamically significant and
the singularities in generic four dimensional
space-times are space-like and
oscillatory  (mixmaster behavior). 3. In the case
of stiff matter, $\gamma=2$, the matter is not insignificant near the
singularity (leading to non-oscillatory behavior)
and generically have anisotropic singularities which
are space-like and non-oscillatory.

We shall investigate the BKL conjectures in cosmological models with an 
effective equation of
state $\gamma = 2$
 and extend them to models with $\gamma > 2$
by calculating the past asymptotic decay rates in general ($G_{0}$)
spatially inhomogeneous models. 
These results
are then supported by a numerical analysis of the behavior of 
spatially inhomogeneous solutions
to Einstein's equations near an initial singularity 
in a special class of
Abelian $G_{2}$ spatially inhomogeneous models.

There are a number of cosmological models of current physical interest which have
 an effective equation of
state $\gamma\ge 2$.
Although a complete fundamental theory is not presently known the
phenomenological consequences can be understood by studying an effective low-energy
theory, which leads to the introduction of additional fields (e.g., scalar fields) in
the high curvature regime close the Planck time scale.  Scalar fields are believed to
be abundant and pervasive in all fundamental theories of physics applicable in the
early Universe, particularly in dimensionally reduced
higher-dimensional theories \cite{Green1987,Olive1990a}. In addition, scalar field cosmological
models are of great importance in the study of the early Universe, particularly in the
investigation of inflation \cite{Olive1990a,inf} and ``quintessence" 
scalar field
models \cite{Caldwell1998a} consistent with observations of the present accelerated
cosmic expansion \cite{PR}.  Superstring theory represents the most promising candidate
for a unified theory of the fundamental interactions, including gravity
\cite{Green1987}.  It is widely believed that eleven-dimensional supergravity
represents the low-energy limit of $M$-theory \cite{Witten95}.  In the low-energy
limit, to lowest-order in both the string coupling and the inverse string tension, all
massive modes in the superstring spectrum decouple and only the massless sectors
remain, which are determined by the corresponding supergravity actions.  A definitive
prediction of string theory is the existence of a scalar field, known as the dilaton,
interpreted as a modulus field parametrizing the radius of the eleventh dimension. There
are two further massless excitations that are common to all five perturbative string
theories, namely the metric tensor field (the graviton) and an anti-symmetric form
field.  When the higher-dimensional metric is compactified, additional form fields and
scalar moduli fields are produced. Thus the low-energy effective action of the theory
essentially reduces to GR plus massless scalar fields.
A massless scalar field (or moduli field etc.) close 
to the initial singularity has an effective equation of state
$\gamma =2$.

Models in which $\gamma\ge 2$ arise naturally in ekpyrotic
\cite{Kho01A} and cyclic \cite{Ste02} cosmological models, which have
a big crunch/big bang transition with a contraction phase dominated by
a scalar field with $\gamma\ge 2$ to the future \cite{design}. In
particular,
it was shown  \cite{erik} that if $\gamma>2$,
chaotic mixmaster
oscillations due to anisotropy and curvature are suppressed and the
contraction is described by a spatially homogeneous and isotropic evolution.
This result was subsequently generalized to theories where the
scalar field couples to $p$-forms, and it 
was also shown that $\mathbb{Z}_2$ orbifold
compactification also contributes to suppressing chaotic behavior. Indeed, 
it was concluded that chaos is avoided in contracting heterotic $M$-theory
models if $\gamma>2$ at the crunch.

There is currently great interest in
higher-dimensional gravity theories inspired by string theory in which the
matter fields are
confined to a $3+1$-dimensional ``brane-world" embedded in higher
dimensions, while the gravitational
field can also propagate in the extra bulk dimensions 
\cite{rubakov}.  There 
has been particular interest in the dynamics of the Universe at early times
in Randall-Sundrum-type brane-world cosmological
models  \cite{randall}.  A unique feature of
brane cosmology is that $\rho^2$ dominates at early times, 
leading to an effective equation of state parameter 
with $\gamma > 2$, which will give rise
to completely different
behavior to that in GR.  
The cosmological implications of brane world
models have been extensively investigated 
\cite{Maartens}.
In particular, it was found that an isotropic singularity is a 
past-attractor in all  orthogonal Bianchi models \cite{Coley}. 
Moreover, the asymptotic dynamical evolution of spatially
inhomogeneous brane-world cosmological models close to the initial
singularity was studied numerically  \cite{CHL} and it was found that there always 
exists an initial singularity,
characterized by the fact that spatial derivatives are
dynamically negligible, which is isotropic 
for all physical parameter values. The numerical results
were supported by a qualitative dynamical analysis and a 
calculation of the past asymptotic decay rates \cite{CHL}.

Andersson and Rendall \cite{Andersson2001} proved that
a generic inhomogeneous cosmology with a stiff fluid or a scalar field
tends to a velocity-dominated solution, which is 
 the Jacobs solution \cite[p. 1426]{WainwrightHsu89}, along individual
timelines.
For a generic inhomogeneous cosmology
with $\gamma >2$, we will show that it tends to
a flat isotropic solution%
\footnote{Namely the
Bin\'etruy, Deffayet and Langlois solution \cite{BDL}, which is
essentially the
flat Friedmann-Lema\^{\i}tre solution \cite[Ch. 2]{WainwrightEllis97}
with $\gamma>2$.}
 along individual timelines. 
Note that these solutions are self-similar \cite{Carr}.
In this paper we shall show that these results 
are supported by the calculations of past asymptotic decay rates
in general $G_0$ models (with $\gamma  \ge2$)%
\footnote{We assume that a massless scalar field (or a massive scalar
field close to the initial singularity) can be modelled by a perfect
fluid with a stiff equation of state. This is justified in detail in
\cite{Andersson2001}.}
and numerical simulations
in a class of $G_{2}$ models with one tilt
degree of freedom.

\section{Asymptotic dynamics at early times for $G_2$ cosmologies}

We first consider the dynamics of a class of $G_{2}$ 
spatially inhomogeneous
cosmological models with one spatial degree of freedom. The governing system
of evolution equations constitute a system of autonomous partial
differential equations in two independent variables. We follow the
formalism of \cite{elst} which utilizes area expansion-normalized
scale-invariant dependent variables,
 and we use the
separable area gauge to consider analytically and numerically the
asymptotic evolution of the class of $G_2$ cosmologies with one tilt
degree of freedom%
\footnote{This class is described by the area expansion rate
$\beta\equiv H-\frac12\sigma_{11}$ \cite[eq. (2.103)]{thesis}
and the normalized variables $(\EEE,A,\Nc,\Nm,\Sigc,\Sigt,
\Sigp,\Sigm,\Omega,v)$ defined with
respect to a time-like congruence that corresponds to the separable area 
gauge.
$\EEE$ is the frame coefficient,
$A,\Nc,\Nm$ are
the components of spatial curvatures, $\Sigc,\Sigt,\Sigp,\Sigm$ are
the shears, $\Omega$ is the normalized density parameter,
and $v$ is the tilt of the perfect fluid.
See \cite[Appendix D]{thesis} for the governing
equations, where we set $\Lambda=0$ and evolve $\beta$ instead \cite[eq.
(2.108)]{thesis}.
We remind the reader that the logarithmic time variable $t\equiv\ln \ell$
is used.}
near the cosmological initial
singularity. For recent works on these models (with $\gamma<2$), we
refer to \cite{thesis}; for vacuum $G_2$ models, we refer to
\cite{Andersson05} and references therein.
The decay rates at early times can be derived by following the analyses in
\cite{Lim04} and \cite{CHL}
by exploiting asymptotic silence.%
\footnote{See \cite{Uggla03} for the notion of the silent boundary and its 
role in past asymptotic dynamics.}
The results for $G_2$ models with one tilt
degree of freedom are given below.

\subsection{Case $\gamma=2$}\label{sec:2A}

For the case $\gamma=2$, we assume that the following conditions hold
uniformly for open intervals of $x$.
\begin{align*}
C_1: &\lim_{t\rightarrow -\infty}
	(\EEE,A,\Nc,\Nm,\Sigc,\Sigt,v)
        = \mathbf{0}, 
	\quad
	\lim_{t\rightarrow -\infty} (\Sigp,\Sigm) =
	(\hatS_+(x),\hatS_-(x))
\\
C_2: &\ \partial_x (\EEE,A,\Nc,\Nm,\Sigc,\Sigt,v, \Sigp,\Sigm)
         \,\,\, \text{are bounded as} \,\,\,  t \rightarrow
-\infty.
\\
C_3: &\  V= {\cal O} (f(t)) \,\,\,\text{implies}\,\,\, \partial_x V =
{\cal O} (f (t)) \,\,\, \text{(asymptotic expansions in time}
\\
        &\ \text{can be differentiated with respect to the
        spatial coordinates)}.
\end{align*}
The decay rates as $t \rightarrow -\infty$ are then given by
\begin{gather}
	(\EEE,A) = e^{2t} [ (\hatEEE,\hat{A}) + \bigO(f)]
\ ,\quad
        (\Sigp,\Sigm,\Omega) = 
	(\hatS_+,\hatS_-,\hat{\Omega}) + \bigO(g)
\\
	\Sigt = e^{k_1 t} [ \hatS_2 + \bigO(g)]
\ ,\quad
        \Sigc = e^{k_2 t} [ \hatS_\times + \bigO(h)]
\ ,\quad
        \Nm = e^{k_3 t} [ \hatN_- + \bigO(h)]
\\
        \Nc = e^{2t} [ - \hatEEE \partial_x \hatS_- \, t + \hatN_\times
			+ \bigO(h)]
\ ,\quad
	v = e^{2t} [ - \tfrac{1}{2} \hatEEE \partial_x \ln \hat{\Omega}
		\, t
		+ \hat{v} + \bigO(h) ]
\end{gather}
where $\hatS_+ := \frac{1}{2}[1-\hatS_-^2 -\hat{\Omega}]$, and
\begin{gather}
	k_1 = -3\hatS_+ + \sqrt{3} \hatS_-,\quad
	k_2 = -2\sqrt{3} \hatS_-,\quad
	k_3 = 2(1+\sqrt{3}\hatS_-),
\\
	f = e^{4t}+e^{2k_1 t},\quad
	g = te^{4t}+e^{2k_1 t}+e^{2k_2 t}+e^{2k_3 t},\quad
	h = t(e^{4t}+e^{2k_1 t}+e^{2k_2 t}+e^{2k_3 t}).
\end{gather}
The area expansion rate $\beta$ and its associated deceleration parameter 
satisfy
\be
        \beta = e^{-3(1-\hatS_+)t} [ \hat{\beta} + \bigO(g) ] 
	\ ,\quad
        q = 2 + \bigO(f)\ .
\ee
The ten hat variables above are functions of $x$, and satisfy the two
constraints
\begin{gather}
	0= \hat{E}_1{}^1 \partial_x \ln \hat{\beta} + \hat{r}\ ,
\quad
	0 = \hat{E}_1{}^1 \partial_x \ln \hatS_2
	-\left[ \hat{r}+3\hat{A}-\sqrt{3}\hat{N}_\times\right]\ ,
\label{G2s_con}
\end{gather}
where
$
	\hat{r} = - 3\hat{A}\hatS_+ - 3\hat{N}_\times\hatS_-
                + 3 \hat{N}_- \hatS_\times -3\hat{v}\hat{\Omega}
$.
That leaves eight hat variables, the same number as when we
specify the initial conditions for numerical simulations.%
\footnote{Only six of the hat variables are essential, since we can use
the remaining temporal gauge freedom to set $\hat{A}=0$, and parameterize
$x$ to set $\hat{E}_1{}^1 =1$.}
To ensure the convergence of $\Nm$, $\Sigc$ and $\Sigt$,
the exponents $k_1$, $k_2$ and $k_3$ must be positive for all $x$.
This implies that the attractor is confined within the
region 
\be
	\Sigm^H > - \frac{1}{\sqrt{3}}(1+\Sigp^H)\ ,\quad
	\Sigm^H <0\ ,\quad
	\Sigm^H > \sqrt{3} \Sigp^H
\ee
inside the Kasner circle
(triangle I in Figure~\ref{new}a) 
 in the state space
of Hubble-normalized variables, where
\be
	\Sig_\pm^H = \frac{\Sig_\pm}{1-\Sigp}\ .
\ee
Note that this restriction is gauge-dependent.

The area expansion-normalized Weyl scalar $\mathcal{W}$
is given by
\be
	\mathcal{W}^2 = \tfrac{1}{9} (\hatS_+ - \hatS_-^2)^2 
			+ \tfrac{1}{9} (1-3\hatS_+)^2\hatS_-^2
			+ \bigO(g+t^2e^{4t})\ .
\ee

\begin{figure}
  \begin{center}
	\epsfig{figure=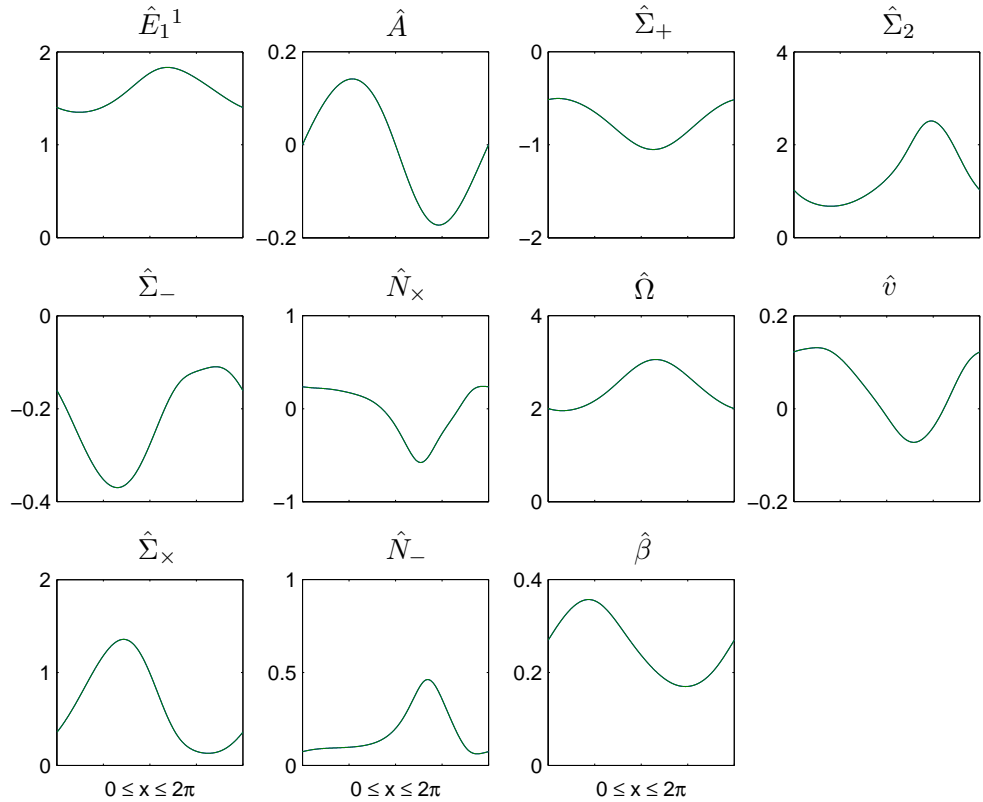,width=13cm}
\setlength{\unitlength}{1mm}
\begin{picture}(120,0)(0,0)
\put(12,112){\makebox(0,0)[l]{$\hat{E}_1{}^1$}}
\put(45,112){\makebox(0,0)[l]{$\hat{A}$}}
\put(78,112){\makebox(0,0)[l]{$\hat{\Sigma}_+$}}
\put(111,112){\makebox(0,0)[l]{$\hat{\Sigma}_2$}}
\put(12,77){\makebox(0,0)[l]{$\hat{\Sigma}_-$}}
\put(45,77){\makebox(0,0)[l]{$\hat{N}_\times$}}
\put(78,77){\makebox(0,0)[l]{$\hat{\Omega}$}}
\put(111,77){\makebox(0,0)[l]{$\hat{v}$}}
\put(12,42){\makebox(0,0)[l]{$\hat{\Sigma}_\times$}}
\put(45,42){\makebox(0,0)[l]{$\hat{N}_-$}}
\put(78,42){\makebox(0,0)[l]{$\hat{\beta}$}}
\end{picture}
\caption{The hat variables computed at $t=-50$ and $t=-100$
coincide, thus confirming the decay rates.}\label{G2s_hat}
  \end{center}
\end{figure}

Figure~\ref{G2s_hat} shows
the hat variables as computed at $t=-50$ and $t=-100$ for a numerical
run (e.g., we plot $e^{-2t}\EEE$, $e^{-2t}(\Nc+t\EEE\partial_x \Sigm)$,
etc).%
\footnote{For numerical simulations, we use {\tt CLAWPACK}. Since it handles
exponential growth rather inaccurately, we evolve the following variables
instead:
\[
        \ln\EEE,\ A/\EEE,\ \Sigm,\ e^{-2t}\Nc,\ \ln\Sigc,\ \ln\Nm,\
        \ln\Omega,\ e^{-2t} {\rm artanh}(v),\ \ln\Sig_2,\ \ln\beta.
\]
Compare with \cite[Appendix D]{thesis}.
One should ensure that $\Sigc$ and $\Nm$ are positive during the
simulation.
We use 32768 grid points with periodic boundary condition, and run the
simulation from $t=0$ to $t=-100$.
For the initial condition, we set $\gamma=2$, $\beta_0=1$ and
\[
        \epsilon=0.1,\ (\EEE)_0=1,\ A_0=0,\ (\Sigm)_0=-0.2,\
        (\Sigc)_0=0.2,\ (\Nm)_0=0.2,\ \Omega_0=1,\ (\Sig_2)_0=0.4
\]
in conjunction with \cite[eq. (9.1)]{thesis}.}
Since the plots at both times coincide, the decay rates are confirmed.
Although the constraints tend to zero, they 
do not tend to zero fast enough and equation 
(\ref{G2s_con}) is not satisfied to a sufficient degree of
accuracy (this numerical problem can be solved by standard methods).
Nonetheless, the numerical results are good enough for confirming the
decay rates.
The numerical simulations consequently confirm the calculations of the decay rates,
provide evidence for the conditions $C_1$--$C_3$, and hence lend support
to the conjectures formulated above.

\subsection{Case $\gamma>2$}

For the case $\gamma>2$, we assume that the following conditions hold
uniformly for open intervals of $x$.
\begin{align*}
C_1: &\lim_{t\rightarrow -\infty}
        (\EEE,A,\Nc,\Nm,\Sigp,\Sigm,\Sigc,\Sigt,v)
        = \mathbf{0},
\\
C_2: &\ \partial_x (\EEE,A,\Nc,\Nm,\Sigc,\Sigt,v, \Sigp,\Sigm)
         \,\,\, \text{are bounded as} \,\,\,  t \rightarrow
-\infty.
\\
C_3: &\  V= {\cal O} (f(t)) \,\,\,\text{implies}\,\,\, \partial_x V =
{\cal O} (f (t)) \,\,\, \text{(asymptotic expansions in time}
\\
        &\ \text{can be differentiated with respect to the
        spatial coordinates)}.
\end{align*}
The decay rates as $t \rightarrow -\infty$ are then given by
\begin{gather}
	(\EEE,A,\Nm,\Nc,v)= \eta \left[
	(\hatEEE,\hat{A},\hatN_-,\hatN_\times,\hat{v}) + \bigO(\xi)\right]
\\
	(\Sigp,\Sigm,\Sigc,\Sigt,\Omega-1)
	= \xi \left[
	(\hatS_+,\hatS_-,\hatS_\times,\hatS_2,-2\hatS_+)
	+\bigO(\xi)\right]
\end{gather}
where
\be
\label{etab}
	\eta = e^{\frac{1}{2}(3\gamma-2)t},\quad
	\xi = e^{\frac{3}{2}(\gamma-2)t}\ .
\ee
The area expansion rate $\beta$ and its associated deceleration parameter
satisfy
\be  
        \beta = e^{-\frac{3}{2}\gamma t} [ \hat{\beta} + \bigO(\xi) ]
        \ ,\quad
        q = \tfrac{1}{2}(3\gamma-2) + \bigO(\xi)\ .
\ee
The ten hat variables above are functions of $x$, and satisfy the two
constraints
\begin{gather}
	0 = \hat{E}_1{}^1 \partial_x \ln \beta 
		-\tfrac{3}{2}\gamma\hat{v} 
\ ,\quad
        0 = \hat{E}_1{}^1 \partial_x \ln \hatS_2
        -( -\tfrac{3}{2}\gamma \hat{v}
        +3\hat{A}-\sqrt{3}\hat{N}_\times )
	\ .
\label{G2b_con}
\end{gather}
That leaves eight hat variables, the same number as when we
specify the initial conditions for numerical simulations.

The area expansion-normalized Weyl scalar $\mathcal{W}$ is given by
\be
	\mathcal{W}^2 = 
	\tfrac{1}{9}(\hatS_+^2 + \hatS_-^2 + \hatS_\times^2 + \hatS_2^2)
	\xi^2 + \bigO(\xi^3)\ .
\ee

\begin{figure}
\begin{center}
	\epsfig{figure=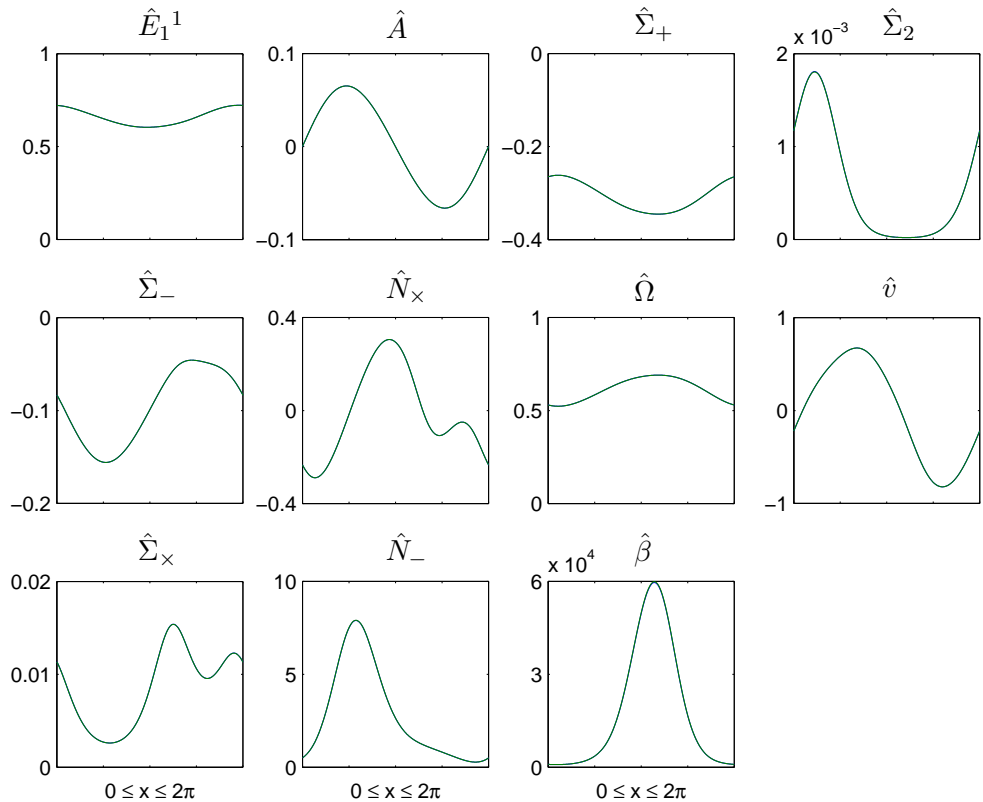,width=13cm}
\setlength{\unitlength}{1mm}
\begin{picture}(120,0)(0,0)
\put(12,112){\makebox(0,0)[l]{$\hat{E}_1{}^1$}}
\put(45,112){\makebox(0,0)[l]{$\hat{A}$}}
\put(78,112){\makebox(0,0)[l]{$\hat{\Sigma}_+$}}
\put(111,112){\makebox(0,0)[l]{$\hat{\Sigma}_2$}}
\put(12,77){\makebox(0,0)[l]{$\hat{\Sigma}_-$}}
\put(45,77){\makebox(0,0)[l]{$\hat{N}_\times$}}
\put(78,77){\makebox(0,0)[l]{$\hat{\Omega}$}}
\put(111,77){\makebox(0,0)[l]{$\hat{v}$}}
\put(12,42){\makebox(0,0)[l]{$\hat{\Sigma}_\times$}}
\put(45,42){\makebox(0,0)[l]{$\hat{N}_-$}}     
\put(78,42){\makebox(0,0)[l]{$\hat{\beta}$}} 
\end{picture}
\caption{The hat variables computed at $t=-50$ and $t=-100$
coincide, thus confirming the decay rates.}\label{G2b_hat}
\end{center}
\end{figure}

Figure~\ref{G2b_hat} shows
the hat variables as computed at $t=-50$ and $t=-100$ for a numerical
run (e.g. plot $\EEE/\eta$, etc).%
\footnote{For the case $\gamma>2$, we evolve the variables
\[
        \ln\EEE,\ A/\EEE,\ \Sigm/\xi,\ \Sigc/\xi,\
        \Nm/\eta,\ \Nc/\eta,\ (\Omega-1)/\xi,\ {\rm artanh}(v)/\eta,\
                \ln\Sig_2,\ \ln\beta.
\]
We use the same initial condition as before, except with
$\gamma=2.1$.}
That the plots at both times coincide confirms the decay rates.%
\footnote{We again note that numerically the constraints do not tend to
zero sufficiently fast.} 
Therefore, we have presented numerical evidence for isotropization in these models.

\section{Asymptotic dynamics at early times for $G_0$ cosmologies}

It would also be of interest to investigate general inhomogeneous ($G_0$) models
with $\gamma \ge 2$ close to the singularity. 
We refer to \cite{Uggla03} for the equations in the separable volume
gauge, using
Hubble-normalized variables (where we set $\Lambda=0$, $R^\alpha=0$).
We give the asymptotic decay rates below. We hope to be
able to numerically simulate the $G_0$ cosmologies in future work.%
\footnote{Simulation of $G_0$ cosmologies is expensive and technically
difficult.
Simulations of vacuum $G_0$ cosmologies with 50 grid points for each of
$x^i$ has been carried out recently \cite{Garfinkle04}.}

\subsection{Case $\gamma=2$}

For the case $\gamma=2$, we assume that the following conditions hold
uniformly for open sets of $x^i$.
\begin{align*}
C_1: &\lim_{t\rightarrow -\infty}
        (E_\alpha{}^i,A_\alpha,N_{\alpha\beta},v_\alpha)
        = \mathbf{0},
        \quad
        \lim_{t\rightarrow -\infty} \Sig_{\alpha\beta} =
		\hatS_{\alpha\beta}
\\
C_2: &\ \partial_x
	(E_\alpha{}^i,A_\alpha,N_{\alpha\beta},v_\alpha,\Sig_{\alpha\beta})
         \,\,\, \text{are bounded as} \,\,\,  t \rightarrow
-\infty.
\\
C_3: &\  V= {\cal O} (f(t)) \,\,\,\text{implies}\,\,\, \partial_x V =
{\cal O} (f (t)) \,\,\, \text{(asymptotic expansions in time}
\\
        &\ \text{can be differentiated with respect to the
        spatial coordinates)}.
\end{align*}
Without loss of generality, we perform a spatially-dependent rotation to
set $\hatS_{\alpha\beta}=0$ for $\alpha\neq\beta$.
The decay rates as $t \rightarrow -\infty$ are then given by (no summation
over repeated indices below)
\begin{gather}
	E_\alpha{}^i = e^{(2-\hatS_{\alpha\alpha})t} 
		[ \hat{E}_\alpha{}^i + \bigO(F) ]
\ ,\quad
        r_\alpha = e^{(2-\hatS_{\alpha\alpha})t}
                [ \hat{r}_\alpha + \bigO(t F) ]
\\
	A_\alpha = e^{(2-\hatS_{\alpha\alpha})t}
		[ \tfrac{1}{2} \hat{E}_\alpha{}^i \partial_i
		\hatS_{\alpha\alpha} \ t + \hat{A}_\alpha 
		+ \bigO(t F) ]
\\
	N_{\alpha\alpha} = e^{(2+2\hatS_{\alpha\alpha})t}
		[ \hat{N}_{\alpha\alpha} + \bigO( t F ) ]
\\       
	N^{\alpha\beta} = e^{(2-\hatS_{\mu\mu})t}
		[ - \hat{E}_\gamma{}^i \partial_i
		\epsilon^{\gamma\delta(\alpha} \hatS{}^{\beta)}{}_\delta
		\ t
		+ \hat{N}^{\alpha\beta} + \bigO( t F ) ]
		\ ,\quad \mu\neq \alpha\neq \beta
\\
	\Omega = \hat{\Omega} + \bigO(F)
\ ,\quad
	v_\alpha = e^{(2-\hatS_{\alpha\alpha})t}
		[ - \tfrac12(\hat{E}_\alpha{}^i \partial_i 
		\ln \hat{\Omega} - 2 \hat{r}_\alpha) \ t
		+ \hat{v}^\alpha + \bigO(t F) ]
\\
	\Sig_{\alpha\alpha} = \hatS_{\alpha\alpha} + \bigO(F)
\ ,\quad
	\Sig_{\alpha\beta} = \bigO(F)
				\ ,\quad \alpha\neq \beta
\end{gather}
where
\[
	F = t^2 e^{2(2-s_+)t} + e^{2(2+2s_-)t},\quad
	s_+(x^i) = \max_{\alpha=1,2,3} \hatS_{\alpha\alpha},\quad
	s_-(x^i) = \min_{\alpha=1,2,3} \hatS_{\alpha\alpha}.
\]
The Hubble scalar and the deceleration parameter satisfy
\be             
        H = e^{-3 t} [ \hat{H} + \bigO(F) ] \ ,\quad
        q = 2 + \bigO(F)\ .
\ee
The twenty eight hat variables above%
\footnote{Note that 
$\hat{\Sigma}_{11}+\hat{\Sigma}_{22}+\hat{\Sigma}_{33}=0$, and we can
write $\hatS_{\alpha\alpha} = \text{diag}(-2\hatS_+,\hatS_+ +
\sqrt{3}\hatS_-,\hatS_+ - \sqrt{3}\hatS_-)$.} 
 are functions of $x^i$, and satisfy the
following sixteen constraints
\begin{gather}
        \hat{r}_\alpha = - \hat{E}_\alpha{}^i \partial_i \ln \hat{H}
\ ,\quad
        0 = \hat{E}_\alpha{}^i \partial_i \hat{\Sig}_{\alpha\alpha}
	+ (2 - \hatS_{\alpha\alpha}) \hat{r}_\alpha
	- 3 \hat{A}_\alpha \hatS_{\alpha\alpha}
-\epsilon_\alpha{}^{\beta\gamma}\hat{N}_{\beta\delta}\hatS_\gamma{}^\delta
	+ 6 \hat{\Omega} \hat{v}_\alpha\ ,
\\
        \hat{\Omega} = 1 - \tfrac{1}{6}
        \left( \hatS_{11}{}^2 + \hatS_{22}{}^2 +
                \hatS_{33}{}^2 \right)\ ,\quad
        0 = 2 (\hat{E}_{[\alpha}{}^j \partial_j - \hat{r}_{[\alpha}
        - \hat{A}_{[\alpha}) \hat{E}_{\beta]}{}^i -
        \epsilon_{\alpha\beta\delta}\hat{N}^{\delta\gamma}
        \hat{E}_\gamma{}^i
        \ .
\end{gather}
That leaves twelve hat variables, of which eight are essential, since we
can
use the remaining temporal gauge freedom to set $\hat{H}=const$ and
make a change of coordinates to set three of the $\hat{E}_\alpha{}^i$'s.

Note that $\hatS_{\alpha\alpha}$ satisfy $|\hatS_{\alpha\alpha}|<2$ by
definition. The exponents of $N_{\alpha\alpha}$ must be positive, however,
thus requiring that $\hatS_{\alpha\alpha}>-1$; i.e.,
the Jacobs solutions are restricted within the triangle
\be
	\Sigp < \frac{1}{2}\ ,\quad
	-\frac{1}{\sqrt{3}}(1+\Sigp) < \Sigm 
		< \frac{1}{\sqrt{3}}(1+\Sigp)\ .
\ee
See triangle I in Figure~\ref{new}b.

The Hubble-normalized Weyl scalar $\mathcal{W}$ is given by
\be
	\mathcal{W}^2 = \tfrac{1}{9}(\hatS_+ + \hatS_+^2 - \hatS_-^2)^2
	+ \tfrac{1}{9}(1-2\hatS_+)^2\hatS_-^2
	+ \bigO(t F)\ .
\ee

The work of Andersson and Rendall \cite{Andersson2001} is
more rigorous, but also assumes more (e.g., analyticity). 
Our results complement their analysis by providing more details on the
spatial curvature variables and on the big O terms, and is easier to apply
and interpret.
We have also discussed the restrictions on the Jacobs disk.

\subsection{Case $\gamma>2$}

For the case $\gamma=2$, we assume that the following conditions hold
uniformly for open sets of $x^i$.
\begin{align*}
C_1: &\lim_{t\rightarrow -\infty}
(E_\alpha{}^i,A_\alpha,N_{\alpha\beta},v_\alpha,\Sig_{\alpha\beta})
        = \mathbf{0},
\\
C_2: &\ \partial_x
(E_\alpha{}^i,A_\alpha,N_{\alpha\beta},v_\alpha,\Sig_{\alpha\beta})
         \,\,\, \text{are bounded as} \,\,\,  t \rightarrow
-\infty.
\\              
C_3: &\  V= {\cal O} (f(t)) \,\,\,\text{implies}\,\,\, \partial_x V =
{\cal O} (f (t)) \,\,\, \text{(asymptotic expansions in time}
\\
        &\ \text{can be differentiated with respect to the
        spatial coordinates)}.
\end{align*}
The decay rates as $t \rightarrow -\infty$ are then given by
\begin{gather}
	(E_\alpha{}^i,\ r_\alpha,\ A_\alpha,\ N_{\alpha\beta},\ v_\alpha)
	= \eta \left[ 
	(\hat{E}_\alpha{}^i,\ \hat{r}_\alpha,\ \hat{A}_\alpha,\
	\hat{N}_{\alpha\beta},\ \hat{v}_\alpha) + \bigO(\xi) \right]
\\
        \Sig_{\alpha\beta} = \xi
                [ \hatS_{\alpha\beta} + \bigO(\xi^2) ]
\ ,\quad
	\Omega = 1 - \tfrac{1}{6}\hatS_{\alpha\beta}\hatS^{\alpha\beta} 
		\xi^2 +\bigO(\xi^4+\eta^2)
\end{gather}
where $\eta$ and $\xi$ are given in (\ref{etab}).
The Hubble scalar and the deceleration parameter satisfy
\be
	H = e^{-\frac{3}{2}\gamma t} [ \hat{H} + \bigO(\xi^2) ] \ ,\quad
	q = \frac{1}{2}(3\gamma-2) + \bigO(\xi^2)\ .
\ee	
The hat variables above are functions of $x^i$, and satisfy the following
constraints
\be
\hat{r}_\alpha = - \hat{E}_\alpha{}^i \partial_i \ln \hat{H},\quad
        0 = 2\hat{r}_\alpha + 3\gamma \hat{v}_\alpha,\quad
        0 = 2 (\hat{E}_{[\alpha}{}^j \partial_j - \hat{r}_{[\alpha}
        - \hat{A}_{[\alpha}) \hat{E}_{\beta]}{}^i -
        \epsilon_{\alpha\beta\delta}\hat{N}^{\delta\gamma}
        \hat{E}_\gamma{}^i
        \ .
\ee

The Hubble-normalized Weyl scalar $\mathcal{W}$ is given by
\be
	\mathcal{W}^2 = \tfrac{1}{9} \cdot \tfrac{1}{6}
	\hatS_{\alpha\beta}\hatS^{\alpha\beta} \xi^2 + \bigO(\xi^3)
	\ .
\ee	

We comment that the results for $G_2$ and $G_0$ cases differ slightly, due
to the difference in both the temporal and spatial gauges employed.

\section{Discussion}

\begin{figure}[h!]
  \begin{center}
      \epsfig{figure=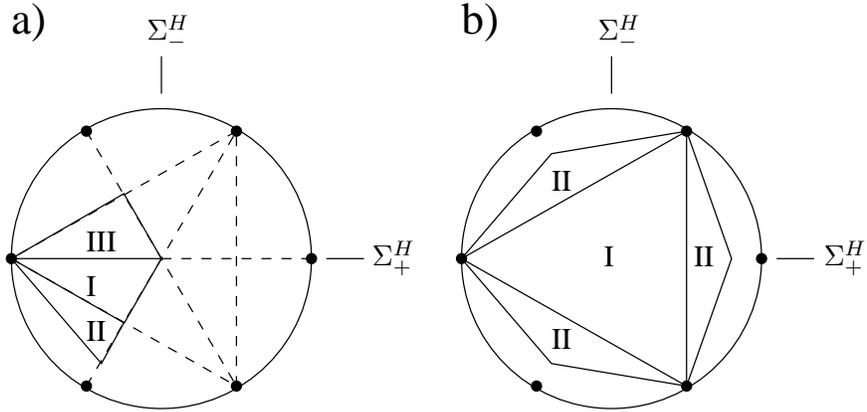, width=12cm}
\setlength{\unitlength}{1mm}
\begin{picture}(120,0)(0,0)
\put(50,32){\makebox(0,0)[l]{$\Sigma_+^H$}}
\put(110,32){\makebox(0,0)[l]{$\Sigma_+^H$}}   
\put(20,62){\makebox(0,0)[l]{$\Sigma_-^H$}}   
\put(80,62){\makebox(0,0)[l]{$\Sigma_-^H$}}
\end{picture}
\caption{a) For the stiff $G_2$ case, the Jacobs solutions that act as the
past attractor are confined within triangle I.
The limits along the conjectured true and fake spike timelines are
confined within triangles II and III
respectively.
b) For the stiff $G_0$ case, the Jacobs solutions that act as the past
attractor are confined within triangle I. The limits along the conjectured
true spike timelines are confined within triangles II.
}\label{new} 
  \end{center}
\end{figure}

We have shown analytically (and confirmed numerically for the $G_2$ models) 
that for the case
$\gamma=2$, a subset of the Jacobs solutions
(triangles I in Figure~\ref{new}) 
 are locally stable into the past, and for the case
$\gamma >2$ the flat Friedmann-Lema\^{\i}tre solution is locally stable
into the past. These results confirm and extend the BKL conjectures, and
complement the work of Andersson and Rendall \cite{Andersson2001}.
They are of also of current physical interest; for example, they lend
support to previous isotropization results \cite{erik,Dunsby}.

In applications to specific physical theories, it may be necessary to
investigate the robustness of these results in the presence of certain
additional fields. 
For example, string and $M$-theories were discussed in order to
motivate the existence of a scalar field (dilaton) in the early universe.
However, in these theories there are also $p$-form fields present, which
can lead to oscillatory behavior close to the cosmological singularity
\cite{Damour00}.

We comment that the conditions $C_1$--$C_3$ in Section~\ref{sec:2A} can
sometimes be violated in a
neighbourhood of isolated timelines $x=const$,
where formation of spiky spatial structures are observed numerically.
In a preliminary numerical investigation for the stiff $G_2$ case, these
structures appear to be similar to
the spikes in vacuum models \cite{Andersson05} and in models with
$\gamma<2$ \cite{thesis}. We conjecture that the limits for $\Sigp$ and
$\Sigm$ along spike timelines reside in the following two triangles: (i)
for true
spikes, the limits reside in the triangle
\be
       - \frac{2}{\sqrt{3}}(1+\Sigp^H) <
        \Sigm^H < -\frac{1}{\sqrt{3}}(1+\Sigp^H)\ ,\quad
        \Sigm^H > \sqrt{3} \Sigp^H\ ,
\ee
(ii) for false spikes, the limits reside in the triangle
\be
        \Sigm^H < \frac{1}{\sqrt{3}}(1+\Sigp^H)\ ,\quad
        \Sigm^H >0\ ,\quad
        \Sigm^H < -\sqrt{3} \Sigp^H
\ee
(see triangles II and III in Figure~\ref{new}a respectively).
We conjecture that
true spikes are physically real, and are reflected by a discontinuous
limit
for the Weyl scalar. On the other hand, false spikes are artifacts of the
rotating frame, and are reflected by a continuous limit for the Weyl
scalar.
It is of interest to study these spikes in more detail.

Similarly, for the stiff $G_0$ case we conjecture that
spikes occur when the limits of $\Sigma_{\alpha\alpha}$ are discontinuous.
When this occurs, the limits are confined to the three triangles
\be
        0 < 4 + 2 \Sig_{\alpha\alpha}-\Sig_{\beta\beta}\ ,\quad
        \hatS_{\alpha\alpha}<-1\ ,
\ee
for all $\alpha \neq \beta$
(see triangles II in Figure~\ref{new}b; also compare with \cite[Figure 
3]{WainwrightHsu89}).

\begin{acknowledgments}
This work was supported by NSERC of Canada. We would like to thank Henk
van Elst, Claes Uggla and Alan Rendall for their helpful comments.
\end{acknowledgments}

\end{document}